\documentclass[doublecol]{epl2} 
\usepackage[utf8]{inputenc}
\usepackage[T1]{fontenc}
\usepackage{multicol, cuted}

\title{Confinement of bosonic and spinning particles in braneworlds}
\shorttitle{Confinement of particles in braneworlds} %Insert here a short version of the title if it exceeds 70 characters

\author{F. E. A. Souza\inst{1} \and G. Alencar\inst{1} \and L. F. F. Freitas\inst{1}  \and R. R. Landim\inst{1}}
\shortauthor{F. E. A. Souza \etal}

\institute{                    
  \inst{1} Departamento de Física, Universidade Federal do Ceará-
Caixa Postal 6030, Campus do Pici, 60455-760, Fortaleza, Ceará, Brazil.%\\
}

\abstract{ In this manuscript we study the confinement of bosonic and spinning test particles in smooth Randall-Sundrum models. For this, we show that it is possible to find an effective potential which describe the motion of the particle over the extra dimension. For the bosonic case it is a known fact that free test particles cannot be localized neither in thin nor thick branes. Recently, a coupling to a scalar field has been used to localize a limited range of masses. Up to now, no mechanism has been found that can trap test particles of any mass over the brane. To solve this, we show that a coupling with the dilaton can trap particles of any mass. Next we analyze the spinning particle. The spin variables $\psi^{P}$ introduces an interaction with the curvature Riemann tensor. This introduces a correction to the effective potential. By analyzing it, we find that the spinning particle can be localized at a position different of the brane. We also show that a spinning particle over the brane can escape to infinity. Therefore, we can conclude that free bosonic and spinning particles are not trapped to the brane.}

\begin{document}

\maketitle

\section{Introduction}
Since the emergence of the Kaluza-Klein models \cite{Kaluza, Klein01, Klein02}, higher dimension gravitational scenarios have attracted attention of physicists. In this context, L. Randall and R. Sundrum (RS) proposed the first braneworld models \cite{RS1, RS2}. Even today, RS models play an important role in the extra dimension context. Mainly, because they allow us recover the well-known  gravitational theory, even with an infinitely large extra dimension. This is possible because gravity is confined on a $4$D hypersurface ($3$-brane) embedded in a $5$D spacetime. Inspired by RS models, many other braneworld models with confined gravity were proposed in different dimensional configuration \cite{Gremm1, Kehagias, Bazeia:2002xg, Bazeia1, Gherghetta, Carlos, Silva, Arkani02, Choudhury, Cendejas}. For all these models, gravity is confined on a $3$-brane, allowing us to recover the $4$D gravitational theory. However, the same is not true for the matter fields, as it was largely verified for all the above models \cite{Bajc:1999mh, Alencar:2014moa, Zhao:2014iqa, Alencar:2015rtc, Alencar:2017dqb, Mendes:2017hmv, Freitas:2018iil, Fu:2015cfa}.

The confinement of the fields in these braneworld leads, evidently, to the localization of the particles related to these fields. However, this interpretation can be obtained only in quantum level. The discussion about dynamics and the confinement of a classical test particle was little discussed in the literature. In fact, dynamics of particles was largely discussed for $5$D Kaluza-Klein models \cite{Gegenberg, Kovacs, Cho:1991ab, Wesson:1999mc, Ponce:2001wf, Ponce:2001ec, Ponce:2002rv}. About confinement, this study was little addressed. For RS models, ref. \cite{Mueck:2000bb} discussed the geodesic motion of `free' test particles in RS-II delta-like braneworld. From this analysis, the authors showed that the motion in the extra dimension $y$ is decoupled of the other dimensions and it is given by 
$$|y(t)|=\frac{1}{2k}\ln\!\left(1-v^{2}k^{2}t^{2}\right).$$
Where $k^{2}>0$ and is related to the cosmological constant. Therefore, ordinary matter $(v^{2}<0)$ is inevitably expelled into the extra dimension and it cannot be confined. A similar discussion was also performed in ref. \cite{Dahia:2007ep} for the thick brane model \cite{Gremm1}. But for this, the authors do not found an analytical solution for $y(t)$. They just found an effective `potential' provided by the curved spacetime and, from this, the confinement was discussed. Just like the delta-like model, the confinement of the `free' particle on this thick brane cannot be attained. Therefore, a localization mechanism to trap the test particle on the $3$-brane is necessary. In Ref. \cite{Dahia:2007ep}, the authors provided the confinement through the coupling of the particle with a scalar field $\phi$. This interaction is made by modifying the `free' particle action through a redefinition of the mass in $5$D given by $M_{0}\to \sqrt{M_{0}^{2}+h^{2}\phi^{2}}$. This mechanism provides the confinement for some values of mass. 

In this context, we will discuss the confinement of a classical test particle in $5$D braneworlds. In doing this, let us propose a new localization mechanism for massive test particles which will allow us the confinement for any value of mass. Beyond this, we will also discuss the confinement of a spinning particle (superparticle) proposed by L. Brink {\it et.al.} in Ref. \cite{Brink}. This model presents some interesting features in $4$D and it was not explored yet in a braneworld context. This paper is organized as follows. First, we discuss the bosonic test particle. In doing this, we will make some comments about the results found in the literature. Next, we present and develop the new localization mechanism. Finally, we conclude with the discussion about the confinement of the spinning particle.

\section{Confinement of Bosonic Test Particle}
In this section, we will discuss the confinement of a bosonic test particle in codimension one braneworlds. In order to get a more general analysis, let us consider a generic background metric given by
\begin{eqnarray}\label{001}
ds^{2}=e^{2A(y)}\eta_{\mu\nu}dx^{\mu}dx^{\nu}+e^{2B(y)}dy^{2}.
\end{eqnarray}
Warp factors $A(y)$ and $B(y)$ are functions only the extra coordinate $y$, which is considered infinitely large. In addition, $\eta_{\mu\nu}$ is the Minkowski metric with signature $(-,+,+,+)$. As mentioned above, this metric is considered just as a background, therefore, we will not consider possible backreaction effects of the particles on it. This generic shape of the metric will allow us to discuss a variety of braneworlds, among them, the delta-like RS-II model \cite{RS1} and also some thick brane models \cite{Gremm1, Kehagias, Bazeia:2002xg, Bazeia1, Landim}.

To start this discussion about the confinement of a bosonic test particle, let us use an alternative form for the action of a free relativistic particle given by
\begin{eqnarray}\label{002}
S=\frac{1}{2}\int\left[E^{-2}g_{PQ}\dot{x}^{P}\dot{x}^{Q}-M^{2}\right]E d\tau.
\end{eqnarray}
In this action, $E$ is the vierbien, $\dot{x}^{P}=\frac{dx^{P}}{d\tau}$ and $M$ is the particle mass in $5$D. For the free case, the particle mass is a constant $M_{0}$. However, as we will also discuss the interacting case (localization mechanism), let us consider that this mass is a function of the extra coordinate $M=M(y)$. Generally, the action used to discuss the dynamics of relativistic particles is that in Eq. (\ref{002a}). However, action (\ref{002}) presents an important advantage to our discussion because it allows us to study the massless case. Just like Eq. (\ref{002a}), action (\ref{002}) is invariant by the arbitrary parameter transformation
$
\tau \to \tau' =f(\tau),
$
since the vierbein $E$ changes as
$
E\to E'=\frac{d\tau}{d\tau'}E.
$
This vierbein plays the role of a tetrad on the worldline of the test particle. However, as we will show below, $E$ is not really a dynamic variable and it can be chosen arbitrarily. The choice of vierbein means setting the parameterization (`gauge' condition).

From the action (\ref{002}), we can obtain the equations of motion
\begin{eqnarray}
\frac{\delta S}{\delta E}=0 \to g_{PQ}\dot{x}^{P}\dot{x}^{Q}+E^{2}M^{2}(y)=0,\hspace{1.35cm} \label{005}\\
\frac{\delta S}{\delta x^{N}}=0 \to \frac{D}{D\tau}\left[E^{-1}\dot{x}^{N}\right]+\delta_{y}^{N}EMM'e^{-2B}=0.\label{006}
\end{eqnarray}

Note that, by replacing the equation (\ref{005}) in the action (\ref{002}), we get
\begin{eqnarray}\label{002a}
S=-\int M(y)\sqrt{-g_{PQ}\dot{x}^{P}\dot{x}^{Q}} d\tau,
\end{eqnarray}
which gets the common action for a test particle when $M(y)=M_{0}$. This show that the actions (\ref{002a}) and (\ref{002}) are equivalent. Below, let us discuss the general procedure for the confinement of the bosonic particle.

\subsection{The General Procedure}

When discussing the localization of a test particle, we must analyze its motion in the extra dimension $y$. If we can associate a potential to the motion in $y$, the particle will be confined when this potential has a minimum located over the brane, namely, $y=0$. In general, the localization of the test particle cannot be achieved for the free case, i.e., $M(y)=M_{0}$ \cite{Mueck:2000bb, Dahia:2007ep}. Therefore, we discuss the general case with arbitrary $M(y)$ to include possible localization mechanisms. 

By using the metric (\ref{001}), the Christoffel symbols are given by
\begin{eqnarray}
\Gamma^{\rho}_{MN}&=&\left[\delta_{N}^{\rho}\delta_{M}^{y}+\delta_{M}^{\rho}\delta_{N}^{y}\right]A',\label{007}\\
\Gamma^{y}_{MN}&=&\delta_{M}^{y}\delta_{N}^{y}B'-\delta_{M}^{\mu}\delta_{N}^{\nu}g_{\mu\nu}e^{-2B}A'.\label{008}
\end{eqnarray}
With this, our equations for $x^P$ in Eq. (\ref{006}) becomes
\begin{eqnarray}
&&\frac{D}{D\tau}\left[E^{-1}\dot{x}^{\mu}\right]=\frac{d}{d\tau}\left[E^{-1}\dot{x}^{\mu}\right]
+2A'E^{-1}\dot{x}^{\mu}\dot{y}=0,\label{006a}\\
&&\frac{D}{D\tau}\left[E^{-1}\dot{y}\right]+EMM'e^{-2B}=\frac{d}{d\tau}\!\left[E^{-1}\dot{y}\right]+E^{-1}B'\dot{y}^{2}\nonumber \\
&&-E^{-1}A'g_{\mu\nu}\dot{x}^{\mu}\dot{x}^{\nu}e^{-2B}+ EMM'e^{-2B}=0.\label{006b}
\end{eqnarray}
In the above equations, {\it prime} means derivative regarding extra coordinate $y$.  

The first step is to choose a `gauge' condition and fix $E$. It is convenient to define it such that the parameter $\tau$ can be identified with the proper-time on the brane. This can be attained by choose $E=e^{2A(y)}$. By using this gauge and equation (\ref{005}) we can show that 
\begin{eqnarray}
g_{\mu\nu}\frac{D}{D\tau}\left[E^{-1}\dot{x}^{\nu}\right]=\frac{d}{d\tau}\left[e^{-2A}g_{\mu\nu}\dot{x}^{\nu}\right]\label{conservA}\\ \dot{y}e^{2B}\left[\frac{D}{D\tau}\left(e^{-2A}\dot{y}\right)+e^{2A}MM'e^{-2B}\right]=\nonumber\\ =\frac{1}{2}\frac{d}{d\tau}\left[e^{2(B-A)}\dot{y}^{2}+M(y)^{2}e^{2A}\right]. \label{conserv} 
\end{eqnarray} 
Therefore, from Eqs. (\ref{conservA}) and (\ref{conserv}) we get 
\begin{eqnarray} 
e^{-2A}g_{\mu\nu}\dot{x}^{\nu}=p_{\mu}=constant,\label{momentum}\\ 
e^{2B(y)-2A(y)}\dot{y}^{2}+M(y)^2e^{2A}=-C^{2}=constant.\label{014} 
\end{eqnarray} 
Where in the first equation we used the definition of the four-momentum 
\begin{eqnarray}\label{defmomentum} 
p_{\mu}=\frac{d\cal{L}}{d\dot{x}^{\mu}}=e^{-2A}g_{\mu\nu}\dot{x}^{\nu}. 
\end{eqnarray} 
In order to interpret the constant $C^{2}$, we can use Eq. (\ref{005}) and (\ref{momentum}), thus, such constant is given by $C^2=p^2$. Therefore, the equation (\ref{014}) can be interpreted as the total energy $E=T+U$ and, from this, we can get an effective `potential'. In doing this, we get \begin{eqnarray}\label{015} 
U_{eff}=M(y)^2e^{2A(y)}-M(0)^2. 
\end{eqnarray} 
Therefore, the localization of massive particles can be taken out just from the analysis of this potential. Below, let us discuss the confinement and the effective mass observed over the brane.

First, the particle will be confined on a brane placed at $y=0$ only if the potential (\ref{015}) satisfies
\begin{equation}\label{018}
U'_{eff}(0)=2M(y)M'(y)e^{2A(y)}(0)=0 \to M'(0)=0
\end{equation}
where we have used the fact that $A'(0)=0$ in braneworld models\cite{Gremm1, Kehagias, Bazeia1, Landim}. Beyond this, by using the above condition it must also satisfy 
\begin{equation}\label{019}
U''_{eff}=2e^{2A(y)}\left[MM''+M^2A''\right]_{y=0}>0.
\end{equation}
We should remember that the above conditions are necessary but not sufficient to confine a particle. If the effective potential has a volcano like shape, a particle with large momentum in extra dimension could escape. Therefore, given that $U_{eff}$ has a maximum at $y_{max}$, the particle will be trapped if $\dot{y}_{max}=0$. Given this, an important question is: what is the observed effective mass of the test particle? This can be determined from Eq. (\ref{014}) and using that $p^2=-m^2_{eff}$. From the above discussion, we can replace $y_{max}$ in Eq. (\ref{014}) with $\dot{y}_{max}=0$ and get
\begin{equation}\label{mass}
m^2_{eff}=M^2(y_{max})e^{2A(y_{max})}.
\end{equation}
Thus, the above expression gives us the effective mass of the particle and it is closely related to the return point $y_{max}$. Evidently, if the return point $y_{max}$ does not exist, the effective particle mass cannot be obtained from (\ref{mass}). For this case, the particle is not confined and the concept of effective mass over the brane is meaningless. Now we can apply the above procedure.

\subsection{The Free Particle}
A simple application of the above conditions is the free particle, with $M=M_0=constant$. For this massive case, condition (\ref{018}) is satisfied and (\ref{019}) imposes the condition
\begin{eqnarray}\label{020}
\left. A''(y)\right\vert_{y=0}>0.
\end{eqnarray}
However, to build the braneworld model, the Einstein's equations provide us with $A''(y)<0$ \cite{RS2, Gremm1, Kehagias, Bazeia:2002xg, Bazeia1, Landim}. In this way, condition (\ref{020}) cannot be satisfied and the localization cannot be attained for any massive test particle. This result is not new and was already discussed in the literature for the delta-like Randall-Sundrum model \cite{Mueck:2000bb} and also in Ref. \cite{Dahia:2007ep} for the thick brane model \cite{Gremm1}. 

However, the authors of Refs. \cite{Mueck:2000bb, Dahia:2007ep} did not consider the massless case. This is because they started from the action (\ref{002a}), which is ill-defined for $M=0$. With our approach, this extension can be obtained trivially. For the massless case, the potential is null and the massless particle can be not be confined.

\section{Localization Mechanisms for Massive Particles}
In this section we must  consider some models to confine test particles.   

\subsection{Coupling With Scalar Field}
As we saw previously, the confinement of free massive particles cannot be attained for RS-like models. To solve this problem, in Ref. \cite{Dahia:2007ep}, the authors provided the confinement of test particles by coupling it with a scalar field $\phi$ in thick brane scenarios. This interaction is introduced in the action (\ref{002}) by modifying the particle mass as 
\begin{equation}\label{100}
M(y)^2=M_{0}^{2}+h^{2}\phi^{2}.
\end{equation}
This mechanism is based on the Yukawa interaction used to provide the spinor mass in field theory. With this mechanism, our conditions (\ref{018}) and (\ref{019}) for confinement can be translated to 
\begin{equation}\label{101}
M_{0}<\sqrt{3}\frac{h}{\kappa},
\end{equation}
where, $\kappa$ is the gravitational coupling constant in $5$D. Thus, this mechanism provides us with an upper limit for the allowed values of confined mass. This upper limit, of course, can be as large as we wish because there is no constraint on the free parameter $h$. 

\subsection{Coupling With Dilaton Field}
It is a know fact that the dilaton field can be used to localize fields in RS-like models. Based on this, in this section we investigate if this kind of coupling can localize a test particle. In Ref. \cite{Kim:2003ch}, the authors discuss the dynamics of particles in a context other than braneworld scenario. In such study, they provide the interaction of particles with the dilaton field by replacing $M_{0}\to M_{0} e^{\lambda\pi}.$
Where $\pi$ is the dilaton. Based on this, let us use this same proposal to discuss the confinement of the massive test particle. For the models where the dilaton is present\cite{Gremm1, Kehagias, Landim}, this field is proportional to the warp factor $A(y)$ and given by $ \pi(y)=-\sqrt{3M_{p}^{3}}A(y),$ where $M_{p}$ is the $5$D Planck scale. Therefore, we get the effective potential 
\begin{equation}\label{102}
U_{eff}=M_0^2\left(e^{2(1-\lambda\sqrt{3M_{p}^{3}})A}-1\right). 
\end{equation}
With this,  the conditions (\ref{018}) and (\ref{019}) will be translated to 
\begin{equation}\label{201}
A'(0)=0,\ \ \ \ -\left[\lambda\sqrt{3M_{p}^{3}}-1\right]A''(0)>0.
\end{equation}
Therefore, since $A'(0)=0$, the first condition is satisfied and as we already discussed for the free case, all the brane models with localized gravity obey $A''(0)<0$. Therefore, the above condition will be satisfied only when
\begin{equation}\label{204}
\lambda>\frac{1}{\sqrt{3M_{p}^{3}}}.
\end{equation}
Note that we did not need to know the exact form of the warp factor $A(y)$, but only its behavior at $y=0$. This result is interesting because, unlike the result (\ref{101}), obtained by using the localization mechanism (\ref{100}), the confinement is obtained for any value of the particle's mass $M^{2}_{0}>0$. In view this, the number of free parameters is reduced and the confinement depends only on the gravitational constant. We also point that $U_{eff}$ in Eq. (\ref{102}) is unbounded from above. Therefore, we get that the dilaton field provides the confining of particles with arbitrary mass.

\section{Spinning Particle}
All the above discussion was performed for a bosonic test particle. Thus, it does not consider any information about the particle's spin. Commonly, `classical description' of spin is carried out in the framework of fields. In fact, there is no classical description of the spin. However, by starting from a suitable Lorentz group representation for classical fields, spin description can be attained after the appropriate quantization process. For the test particle case, spin degree of freedom can be included by using Grassmann variables. These models are called pseudomechanic or superparticle model. This was performed in $4$D for flat \cite{Brink, Berezin} and also curved \cite{Ali} spacetimes, however, this was not discussed yet in braneworlds context. Based on these references, we will discuss below the confinement of spinning particles in codimension one braneworld. In doing this, let us start from the action
\begin{eqnarray}\label{401}
S=\frac{1}{2}\int\left[E^{-1}g_{PQ}\dot{x}^{P}\dot{x}^{Q}-EM^{2}_{0}-ig_{PQ}\psi^{P}\frac{D\psi^{Q}}{D\tau}\right.\nonumber\\
\left.-i\xi\dot{\xi}-i\chi\left(E^{-1}g_{PQ}\dot{x}^{P}\psi^{Q}+M_{0}\xi\right)\right]d\tau.
\end{eqnarray}
Where,
$$\frac{D\psi^{Q}}{D\tau}=\dot{\psi}^{Q}+\Gamma^{Q}_{SR}\psi^{S}\dot{x}^{R}$$
and this covariant derivative is necessary because, unlike $x^{N}$, variable $\psi^{N}$ is a vector by general coordinate transformations. Here, evidently, this action deserves some explications. First, coordinates $x^{\mu},\ y$ and $E$ are commuting variables (c-numbers), and $\psi^{N},\ \xi$ and $\chi$ are anticommuting variables (a-numbers). In this configuration, variables $x^{\mu}$ and $y$ describe the geodesic motion, $E$ is the vierbein, quantities $\psi^{N}$ and $\xi$ will be related to the spin description and $\chi$ is necessary by symmetry requirement. Just like the action (\ref{002}), the action presented above is invariant by reparametrization $\tau\to\tau'=f(\tau)$. Beyond this, action (\ref{401}) is also invariant by the local SUSY transformations,
\begin{eqnarray}
\delta x^{N}=i\alpha \psi^{N},\ \ \ \ \delta E=i\alpha \chi,\ \ \ \ \delta \chi=2\dot{\alpha},\nonumber\\
\delta \xi =M_{0}\alpha+\frac{i}{M_{0}E}\alpha\xi\left(\dot{\xi}-\frac{1}{2}M_{0}\chi\right),\nonumber\\
\delta \psi^{N}=\alpha E^{-1}\left(\dot{x}^{N}-\frac{i}{2}\chi\psi^{N}\right).\nonumber
\end{eqnarray}
Where $\alpha$ is a real anticommuting parameter. For a more detailed discussion about this pseudomechanic, we recommend \cite{Brink, Berezin}. Next, we discuss the equation of motion for the action (\ref{401}). In doing this, our objective is to verify if the new spin variables can modify the effective potential $U_{eff}(y)$ in order to provide the confinement.

From action (\ref{401}), equations of motion can be obtained, and they are given by
\begin{eqnarray}
\frac{\delta S}{\delta E} &\to& g_{PQ}\left[\dot{x}^{P}\dot{x}^{Q}-i\chi\dot{x}^{P}\psi^{Q}\right]+E^{2}M^{2}_{0}=0,\label{402}\\
\frac{\delta S}{\delta \chi} &\to & g_{PQ}\dot{x}^{P}\psi^{Q}+EM_{0}\xi=0,\label{404}\\
\frac{\delta S}{\delta \xi}&\to & \dot{\xi}-\frac{1}{2}M_{0}\chi=0,\label{405}\\
\frac{\delta S}{\delta \psi^{P}} &\to& \frac{D\psi^{P}}{D\tau}-\frac{1}{2}E^{-1}\chi\dot{x}^{P}=0,\label{403}\\
\frac{\delta S}{\delta x^{N}} &\to& \frac{D}{D\tau}\left[E^{-1}g_{PN}\dot{x}^{P}\right]-\frac{i}{2}\frac{d}{d\tau}\left[\chi E^{-1}g_{PN}\psi^{P}\right]\nonumber\\& &+\frac{i}{2}R_{NSQR}\psi^{Q}\psi^{R}\dot{x}^{S}=0.\label{406}
\end{eqnarray}

Before to discuss the confinement of the spinning particle, it is convenient to choose `gauge' conditions for the two symmetries mentioned above. Let us choose $E=e^{2A}$, as in the bosonic particle and $\chi=0$ \cite{Brink}. Therefore, the above equations can be simplified as
\begin{eqnarray}
g_{PQ}\dot{x}^{P}\dot{x}^{Q}+M_0^2 e^{4A}=0,\label{407}\\
g_{PQ}\dot{x}^{P}\psi^{Q}+M_0e^{2A}\xi=0,\label{408}\\
\frac{D\psi^{P}}{D\tau}=0,\label{409}\\
\frac{D}{D\tau}\left[e^{-2A}\dot{x}^{N}\right]+\frac{i}{2}R^{N}_{\ SQR}\psi^{Q}\psi^{R}\dot{x}^{S}=0.\label{410}
\end{eqnarray}
Where, variable $\xi$ is a constant. Here, we can briefly discuss the `meaning' of the above equations. In fact, it is not possible to understand the variables $\psi^{Q}$ or $\xi$ classically. However, after quantization is performed, variables $\psi^{Q}$ and $\xi$ can be identified with Dirac's gamma matrix \cite{Brink}. In this way, the word `meaning' here is related to this interpretation in quantum level after quantization. Equations (\ref{408}) and (\ref{409}) will describe the spin dynamics, with Eq. (\ref{408}) playing the role of the Dirac equation. Unlike the bosonic particle case discussed previously, now geodesic equation (\ref{410}) presents a coupling between the spin and the curvature Riemann tensor. In fact, Eqs. (\ref{408})-(\ref{410}) would be analogous to the Mathisson-Papapetrou equations which describe a spinning body in a curved spacetime \cite{Mathisson, Papapetrou}. Below, let us discuss the confinement.

\subsection{Confinement of Spinning Particle}
As we saw for the case of bosonic particles, we are interested in analyze equations for the component $y$. From the equation (\ref{410}), we get the equation for $x^{M}$ given by
\begin{eqnarray}\label{411}
\frac{D}{D\tau}\left[e^{-2A}\dot{y}\right]+\frac{i}{2}R^{y}_{\ \mu\lambda\rho}\psi^{\lambda}\psi^{\rho}\dot{x}^{\mu}\nonumber\\+iR^{y}_{\ \mu y\rho}\psi^{y}\psi^{\rho}\dot{x}^{\mu}=0,\\
g_{\mu\nu}\frac{D}{D\tau}\left[e^{-2A}\dot{x}^{\nu}\right]+iR_{\mu y y\rho}\psi^{y}\psi^{\rho}\dot{y}\nonumber\\+\frac{i}{2}R_{\mu \nu \rho\alpha}\psi^{\rho}\psi^{\alpha}\dot{x}^{\nu}=0. \label{mu}
\end{eqnarray}
By using the Christoffel symbols (\ref{007}) and (\ref{008}), the components of the Riemann tensor are given by
\begin{eqnarray}
R^{\rho}_{\ SQR}=\delta_{S}^{y}\left[\delta_{R}^{\rho}\delta_{Q}^{y}-\delta_{Q}^{\rho}\delta_{R}^{y}\right]\left[A''+A'^{2}-A'B'\right]\nonumber\\+\left[\delta_{R}^{\rho}\delta_{Q}^{\mu}-\delta_{Q}^{\rho}\delta_{R}^{\mu}\right]\delta_{S}^{\nu}g_{\mu\nu}e^{-2B}A'^{2},\label{412}\\
R^{y}_{\ SQR}=\delta_{S}^{\nu}\!\left[\delta_{R}^{y}\delta_{Q}^{\mu}-\delta_{Q}^{y}\delta_{R}^{\mu}\right]\!\left[A''\!+\!A'^{2}\!-\!A'B'\right]\!g_{\mu\nu}\!e^{-2B}.\nonumber\\\label{413}
\end{eqnarray}
And we get
\begin{eqnarray}
\frac{D}{D\tau}\left[e^{-2A}\dot{x}^{\rho}\right]+i\left[A''-A'B'\right]\psi^{y}\psi^{\rho}\dot{y}\nonumber\\
+iM_0\psi^{\rho}\xi e^{2(A-B)}A'^{2}=0,\label{414a}\\
\frac{D}{D\tau}\left[e^{-2A}\dot{y}\right]=iM_{0}\left[A''+A'^{2}-A'B'\right]e^{2(A-B)}\xi\psi^{y}.\label{414b}
\end{eqnarray}
Where, we already used the constraint (\ref{408}) to eliminate the dependence in $\dot{x}^{\mu}$. To find the constants of motion, we need to use the equations for $\psi^{\rho}$. 

From Eq. (\ref{409}) and by using the Christoffel symbols (\ref{007}) and (\ref{008}), we get 
\begin{eqnarray}
\dot{\psi}^{\rho}+\left[\psi^{y}\dot{x}^{\rho}+\psi^{\rho}\dot{y}\right]A'=0,\label{415}\\
\dot{\psi}^{y}+\psi^{y}(\dot{B}+\dot{A})+\xi M_{0}e^{2A-2B}A'=0.\label{419}
\end{eqnarray}
We can multiply Eq.(\ref{419}) by $\xi$ and use the constraint (\ref{408}) to find the solution
\begin{eqnarray}
\psi^{y}(\tau)\xi=\psi^{y}_{0}\xi e^{-A-B}.\label{421}
\end{eqnarray}
Now, we can multiply Eqs. (\ref{415})  and (\ref{419}) by $\psi^{y}$ and $\psi^{\rho}$, respectively, to get
 \begin{eqnarray}
\frac{d}{d\tau}\left[\psi^{y}\psi^{\rho}\right]+\psi^{y}\psi^{\rho}\left(\dot{B}+2\dot{A}\right)+M_0\xi e^{2A-2B}A'\psi^{\rho}=0.
\end{eqnarray}
Finally, from the above equation, we can show that 
\begin{eqnarray}
\frac{d}{d\tau}\left[\psi^{y}\psi^{\rho}e^{2A+B}\right]+M_0\xi e^{4A-B}A'\psi^{\rho}=0.\label{421b}
\end{eqnarray}

With this expression, we can discuss if the spinning superparticle can be confined. If we use our previous relations  (\ref{conservA}), (\ref{conserv}), (\ref{421}) and (\ref{421b}) in equation (\ref{414a}), we get
\begin{eqnarray}
\frac{d}{d\tau}\left[\dot{x}^{\mu}+i\psi^{y}\psi^{\mu}e^{2A}A'\right]=\frac{dp^{\mu}}{d\tau}=0,\label{conservmod-a}\\
\frac{d}{d\tau}\left[e^{2B-2A}\dot{y}^{2}+M_{0}^{2}e^{2A}-2iM_{0}\xi\psi^{y}_{0}e^{A-B}A'\right]=0. \label{conservmod-b}
\end{eqnarray}
Where, we define the conserved momentum
\begin{equation}\label{supermomentum}
p^{\mu}=\dot{x}^{\mu}+i\psi^{y}\psi^{\mu}e^{2A}A'
\end{equation}
I order to interpret the second constant coming from the second equation in (\ref{conservmod-b}), we must use our mass constraint and equation (\ref{supermomentum}) to get
\begin{eqnarray}
-p^2=e^{2B-2A}\dot{y}^{2}+M_{0}^{2}e^{2A}-2iM_{0}\xi\psi^{y}_{0}e^{A-B}A',\label{424a}
\end{eqnarray}
with $p^2=\eta_{\mu\nu} p^{\mu}p^{\nu}$ and we already used Eq. (\ref{407}). Thus, just as before, we can interpret the term on the right as a total energy and 
\begin{eqnarray}
U_{eff}=M_{0}^{2}\left[e^{2A}-\frac{2i}{M_{0}}\xi\psi^{y}_{0} e^{A-B}A'\right].\label{424}
\end{eqnarray}
First, when $\psi^{N}$ is zero, the last term in right-hand of above equation is zero. Therefore, we get the results discussed previously for a bosonic test particle and given by Eq. (\ref{015}) with $M(y)=M_{0}$. Now, because the spin variable, we get a new peace in $U_{eff}$ acting on the particle.  Thus, the spin modify the effective potential, due to the interaction of it with the curvature. With this we have the possibility of confining a spinning  particle. 

\subsection{Application}
\begin{figure}[!ht]
\centering
\includegraphics[scale=0.55]{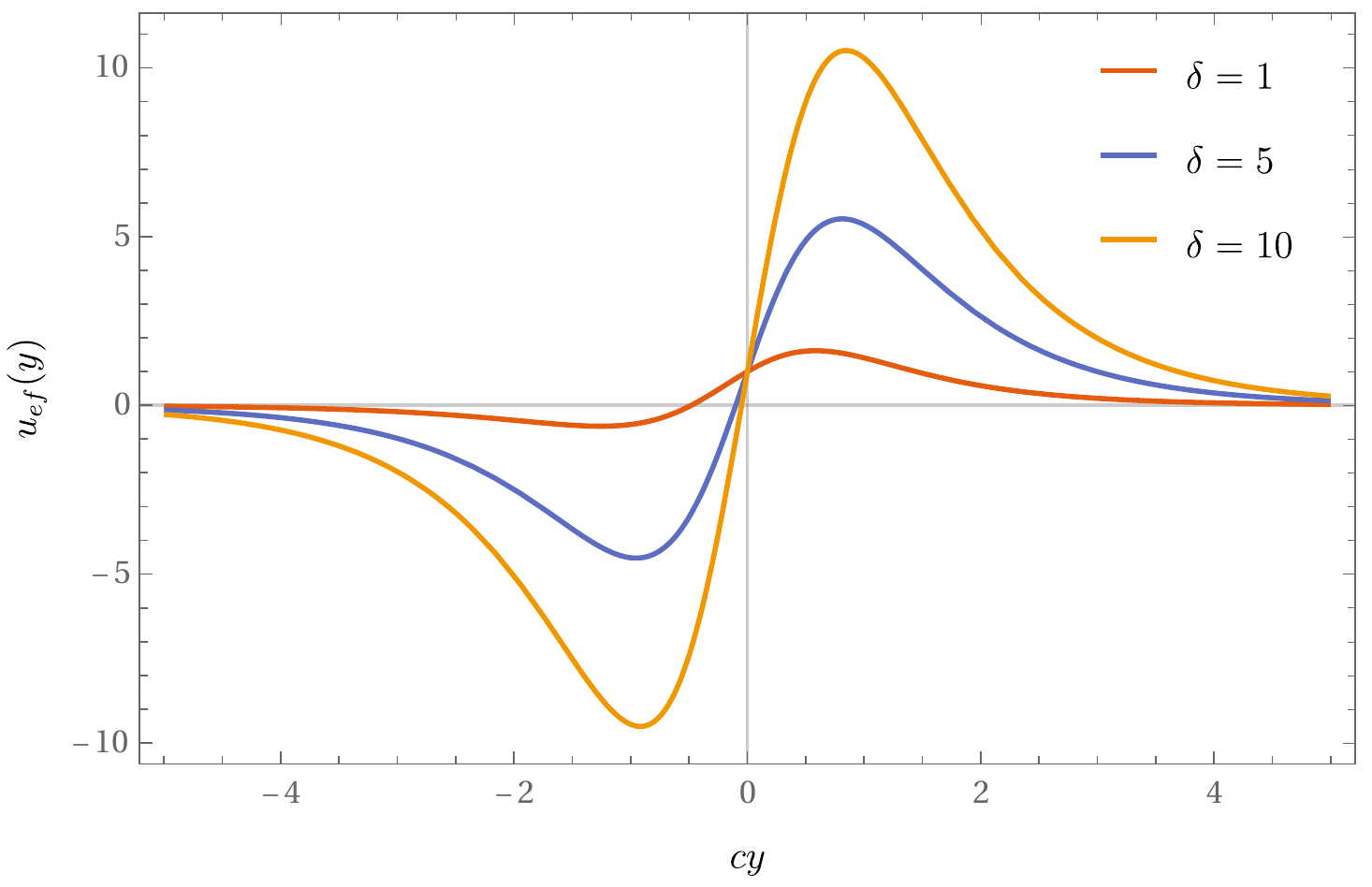}
\caption{Plot of $u_{eff}=U_{eff}/M_{0}^{2}$ for the model (\ref{425}) with $b=1$ and $\delta\equiv\frac{i}{M_{0}}\xi\psi^{y}_{0}$.}\label{fig1}
\end{figure}

For the massless case, we see from Eq. (\ref{424}) that $U_{eff}=0$. Therefore, just as in the bosonic case, the spinning particle is not confined. However, for the massive case $U_{eff}$ has a correction due to the Grassmann variables. Figure (\ref{fig1}) shows a plot of the effective potential (\ref{424}) for the thick brane presented in Ref. \cite{Gremm1}, namely,
\begin{eqnarray}
A(y)=\ln\left[\mbox{sech}^{b}\left(cy\right)\right],\ \ \ \ B(y)=0.\label{425}
\end{eqnarray} 

As we can see, $y=0$ is not an extreme for this effective potential. But, there are two extreme points, and one of them is a minimum. Therefore, the spinning particle can be localized at the minimum of this potential. However, this localization does not happen at $y=0$, but at a point that is slightly displaced from the origin. Similar conclusions are also obtained for other models as, for example, the delta-like brane \cite{RS2}, the thick brane presented in Ref. \cite{Kehagias} or the thick brane with inner structure obtained in Refs. \cite{Bazeia1, Bazeia2003, Landim}. In fact, from Fig. (\ref{fig1}) we can conclude that any particle over the brane will escape to the extra dimension. Note that $U_{eff}(y=0)>0$ and for large $y$ the potential is null. Therefore, any particle over the brane goes to infinity.

An interesting possibility is to look for models in which the minimum of $U_{eff}$ coincides with the position of the brane. Since our $U_{eff}$ is asymmetric we should look for brane models with the same property. They are called asymmetric braneworlds and were discussed by D. Bazeia {\it et.al.} in Refs. \cite{Bazeia2013, Bazeia2019, Bazeia2020}. However, for these models we also discovered that the spinning particle is not localized. It seems that there is some general property of the potential (\ref{424}) that forbids the confining. Let us now show this. First note that  
\begin{eqnarray}
U'_{eff}=2M_{0}^{2}e^{2A}A'-2M_{0}^{2}\delta e^{A-B}\left[A''+A'^{2}-A'B'\right].
\end{eqnarray}
Now we should remember that for any braneworld, we must have that at the position of the brane $A'(y_B)=B'(y_B)=0,A''(y_B)>0$. With this we get that
\begin{eqnarray}
U'_{eff}(y_B)=-2\delta e^{A-B}A''\neq 0
\end{eqnarray}
and therefore at the position of the brane we never have a minimum of $U_{eff}$. However, we could have the possibility that the particle is trapped around the position of the brane and do not escape to infinity. In order to analyze this we use the general properties $A(y_B)=0,e^A(\infty)=0$ to obtain  
\begin{eqnarray}
\Delta U_{eff}\equiv U_{eff}(\infty)-U_{eff}(y_B)=-M_{0}^{2}.
\end{eqnarray}
Therefore the potential at infinity is always fewer than the potential over the brane and any spinning particle over the brane will escape to infinity. 

\section{Conclusion}
In this manuscript, we discussed the confinement of test particles in codimension one braneworlds. We discussed the bosonic test particle and carried out the discussion about confinement of the spinning particle (superparticle). Localization of bosonic test particles was already studied previously in the literature. For the delta-like RS-II model, Ref. \cite{Mueck:2000bb} discussed the confinement of a free test particle. And also, for the thick brane presented in Ref. \cite{Gremm1}, where it was studied the free case and also with a localization mechanism \cite{Dahia:2007ep}.

In discussion of the bosonic test particle, we studied the massive and massless case. For the free massive case, we obtained the conserved quantities given in equations (\ref{momentum}) and (\ref{014}). From this, we showed that the motion in extra dimension can be decoupled of other coordinates and it was possible to identify the effective potential (\ref{015}), namely, 
\begin{eqnarray}\label{300} 
U_{eff}=M(y)^2e^{2A(y)}-M(0)^2. 
\end{eqnarray} 
With this potential, the confinement would be attained if the conditions $U'_{eff}(0)=0$ and $U''_{eff}(0)>0$ could be satisfied. For free massive particles these conditions imply that $A''(y)>0$. However, for all the braneworld model with localized gravity \cite{RS2, Gremm1, Kehagias, Bazeia1, Bazeia2003, Landim}, the Einstein's equations provide us with $A''(y)<0$. Therefore, the necessary condition to confine the massive particles, namely, $A''(y)>0$, cannot be satisfied. This result agrees with the one obtained in Refs. \cite{Mueck:2000bb,Dahia:2007ep} by other means. However, the massless case has not been considered in Refs. \cite{Mueck:2000bb,Dahia:2007ep}, since their actions is ill-defined for $M=0$. By using a quadratic action, we show that (\ref{300}) is valid for both cases. For $M=0$, we get that (\ref{300}) is zero and, thus, the above approach is not conclusive. However, as discussed from Eq. (\ref{021}), the massless particle cannot be localized over the brane.

Next, we proposed a new localization mechanism for the massive particle by replacing 
$$M_{0}\to M(y)=M_{0}e^{\lambda\pi(y)},$$
 where, $\pi(y)$ is a scalar field sometimes called dilaton. This new mechanism is an alternative to that presented in Ref. \cite{Dahia:2007ep}, where it was provided a coupling with a scalar field given by $M_{0}\to \sqrt{M_{0}^{2}+h^{2}\phi^{2}}.$ With this mechanism, the confinement can be attained for particles with mass values that satisfy $M_{0}<\sqrt{3}\frac{h}{\kappa}.$ As we showed, by using our localization mechanism, it was possible to obtain a modified effective potential, but now given by 
\begin{eqnarray}\label{301} 
U_{eff}=M_0^2\left(e^{2(1-\lambda\sqrt{3M_{p}^{3}})A}-1\right). 
\end{eqnarray} 
Once again, we apply the confinement conditions to obtain 
$$\lambda>\frac{1}{\sqrt{3M_{p}^{3}}}.$$
This result presents the interesting feature that it allows us to confine any value of mass $(M_{0}^{2}>0)$. In this sense, the coupling with the dilaton field seems to provide a more efficient confinement of the massive particles.

To conclude the analysis, we discussed the spinning particle presented for $4$D in \cite{Brink, Berezin, Ali}. This study was not yet performed in the context of braneworld. Just like the bosonic test particle, it was discussed the massive and the massless cases. We found the modified effective potential given by (\ref{424}), namely, 
$$U_{eff}=M_{0}^{2}\left[e^{2A}-\frac{2i}{M_{0}}\xi\psi^{y}_{0} e^{A-B}A'\right].$$
Modification in this potential is closely related the presence of the spin variables $\psi^{P}$ and $\xi$, which introduces an interaction of it with the curvature Riemann tensor. The analysis of the confinement was performed graphically. From figures (\ref{fig1}), we concluded that the above potential allows us to localize the test particles with spin in a position close of the brane, but not over the brane. By analyzing the general features of the above potential we show that in fact any particle over the brane will escape to infinity. This is valid for a general warp factors $A(y)$ and $B(y)$. Beyond this, the above potential is also valid for the massless case, and we conclude that they are not localized. Therefore, neither the bosonic nor the spinning free particles can be localized over the brane. This is very different of the field description, where the free scalar field are localized, but spinor and gauge fields are not.

Finally, we should point that due to the presence of worldline supersymmetry the possibility of couplings is very constrained. For example, it is not possible to introduce a coupling with the scalar or dilaton fields above without breaking this symmetry. However, we can consider the case with more than one supersymmetry and even higher spins from the worldline viewpoint \cite{vanHolten:1995qt,Bastianelli:2008nm}. Another possibility is to consider spacetime supersymmetry from the wordline viewpoint, such as the Brink-Schwarz superparticle \cite{Berkovits:2001rb} or even the covariant pure spinor description \cite{Brink:1981nb,Berkovits:2001rb}. Therefore, the introduction of Grassmann variables opens a new avenue of research about confining of test particles in braneworlds.

\acknowledgments
The authors would like to thank Alexandra Elbakyan and sci-hub, for removing all barriers in the way of science. We acknowledge the financial support provided by the Conselho Nacional de Desenvolvimento Científico e Tecnológico (CNPq) and Fundação Cearense de Apoio ao Desenvolvimento Científico e Tecnológico (FUNCAP) through PRONEM PNE-0112-00085.01.00/16.

\end{document}